\documentstyle[epsf]{article}
\begin{document}

\title{Microwave Background Fluctuations in an Open Universe }

\author{David N. Spergel, Ue-Li Pen \\
{\it  Princeton University Observatory, Princeton, NJ 08544 USA} \\
    \\
Marc Kamionkowski \\  
{\it  Institute for Advanced Study, Princeton, NJ 08540 USA} \\
  \\
and \\
  \\
Naoshi Sugiyama\footnote{also,
Department of Physics, Faculty of Science
University of Tokyo,
Tokyo 113, Japan} \\
{\it Dept. of Astronomy,
University of California, Berkeley, CA 94720 USA}}
\maketitle

\section*{Abstract}

There are several models for generating fluctuations in an open
universe that are compatible with the microwave background
fluctuations detected by COBE {\it and} observations of large scale
structure.  Topological defects, such as strings and textures,
appear to be more successful in a low-$\Omega$ model than in an
$\Omega = 1$ model. These models predict that the cosmic
microwave background (CMB) fluctuations
are non-Gaussian with distinctive signatures that may lead to their
confirmation. If there exists a mechanism for generating scale
invariant mass fluctuations, $(\delta M/M)^2 \propto M^{-4/3}$,
then these models are also compatible with COBE. Both models
predict mass fluctuations on the 8/h Mpc scale, $\sigma_8$, $\sim
0.5 - 1$ for $\Omega h \sim 0.2 - 0.3$.
A variety of observations of
large-scale structure, clusters and galaxies, as well as the age
problem, suggest that the low-$\Omega$ models are attractive.

\section{Introduction}

With the notable exception of literally a handful of
papers (e.g., Wilson 1983; Abbott and Schaeffer
1986, Sugiyama and Gouda 1992), most cosmologists
have focused on microwave fluctuations
in a flat Universe.  Concordance of high values for the
Hubble parameter (Jacoby et al. 1992) with conservative lower
bounds on the age of the Universe and
numerous dynamical measures suggesting that $\Omega < 1$
(see White 1992 for review) have encouraged us
to explore open-universe models in more detail.

	Open universes differ {}from flat universes in two distinct ways:
(1) the geometry of an open universe is hyperbolic rather than flat;
and (2) density fluctuations in an open universe stop growing at
redshifts much smaller than $1/\Omega$.  In this lecture, I will
discuss how these two differences lead
to a distinctive microwave background signature for
open-universe models.

\section{Types of Fluctuations:}

The predictions of a cosmological model depend not only on its assumptions
about the global geometry of the Universe, but also on its assumptions
about the mechanism for generating density fluctuations and its assumptions
about the nature of the dark matter.   
Density fluctuations can be divided into two broad categories: curvature
fluctuations and equation-of-state fluctuations.  These two categories
are generalizations of the usual division of fluctuations in 
models composed of only baryons and photons into adiabatic fluctuations
and isothermal fluctuations.  

	Curvature fluctuations are fluctuations in 
the total density of matter and assume no variation in the local properties
of matter.  In inflationary models, quantum fluctuations in the inflaton field
develop into curvature fluctuations and the study of the evolution of these
fluctuations has dominated theoretical cosmology for the past decade.  

Equation-of-state fluctuations are fluctuations that are produced through
local variations in the properties of matter.  In isothermal baryon models,
variations in the ratio of baryons to photons (entropy fluctuations) are the
equation-of-state fluctuations that produce density fluctuations.  In 
topological-defect models, such as cosmic strings and textures, the 
stress-energy in the defects are the source of the variations in equations
of state that lead to density fluctuations.  If there was a primordial
magnetic field, it would produce density fluctuations through variations
in the ratio of magnetic-field energy to the local density.

	The difference between these two types of fluctuations are apparent
when we consider the evolution equation for potential fluctuations in
a flat matter-dominated universe:
\begin{equation}\Phi'' + 2 (a'/a)\Phi' = S\end{equation}
Here, $\Phi$ is  the gravitational potential, $a$
is the scale factor, prime denotes derivatives with respect to conformal
time and $S$ depends on variations in the local equation of state.
In a universe filled only with matter or in a universe with
only adiabatic fluctuations, $S = 0$.  In models
with curvature fluctuations, $S = 0$, and $\Phi$ has an initial
non-zero value at the start of the Friedman-Robertson-Walker phase.
Thus, these models start the FRW phase with variations in the local
curvature.  In models with equation-of-state fluctuations, $S$ is non-zero
and $\Phi(t=0) = 0$.  Thus, in these models, the
universe starts its FRW phase without any variations in curvature.
Hence, these models are also sometimes called isocurvature models. 

\section{Curvature Fluctuations:}

	We will first consider the evolution of 
curvature fluctuations in  an open universe.  In an open universe,
the evolution equation for potential fluctuations is modified (Kodama
and Sasaki 1984):
\begin{equation}
{\cal A}' + {a \over a'} {\cal A} = {a \over a'} S,
\end{equation}
where ${\cal A} = - (a^3 \Phi/a')'/a^2$.
This modification implies that at late times, potential fluctuation
in an open universe decay.  The fluctuation decay has two significant
effects on the microwave background: (a) the
decaying potential fluctuations are an additional source
of CMB fluctuations and (b) since we normalize theories to observations
of structure at the current redshift, this decay implies that open
universe models started out with larger potential fluctuations
at recombination.  Since both of these factors increase the predicted
level of CMB fluctuations, many cosmologists expected that open
universe models were inconsistent with the level of fluctuations
observed by COBE.  However, the amplitude of fluctuations
decreases for larger Hubble constants, and geometric effects
lower the level of fluctuations predicted by open-universe
models on the largest scales, so these models can be compatible
with COBE.

	In a flat universe, an angular cone of radius $10^o$ in radius
subtends  a surface of comoving radius 300 Mpc at the surface of last
scatter.  Thus, if the Universe is flat, then  COBE is probing density
fluctuations on this scale.  However, in an open universe, the same
cone subtends a surface of somewhat larger radius, $\sim 300/\Omega$ Mpc.
Since density fluctuations decrease with scale in almost all proposed
cosmological models, this implies that this geometric effect suppresses
CMB fluctuations.

 Kamionkowski and Spergel (1993) recently completed an analysis of CMB
fluctuations in an open universe that included both geometric and
evolutionary effects.  This analysis assumed that the Universe
was cold-dark-matter dominated and the spectrum of density
fluctuations was scale-invariant.  Figures 1 and 2 show
that the COBE predictions were insensitive to the details of
how we extrapolated the scale-invariant spectrum to scales larger
than the curvature scale.    
Figure 1 shows three different procedures for extrapolating
a scale-invariant power spectrum beyond the curvature scale.
Figure 2 shows the predicted amplitude of the CMB multipoles
for these three very different power law extrapolations.  Since
the COBE experiment is sensitive mostly to the amplitude of
mass fluctuations on the 1000-Mpc scale, it is not very sensitive
to the variations in the form of the power spectrum on superhorizon scales.
The differences between these different extrapolations would be
swamped by cosmic variance.

\begin{figure}[htb]
\epsfxsize=3in \epsfbox{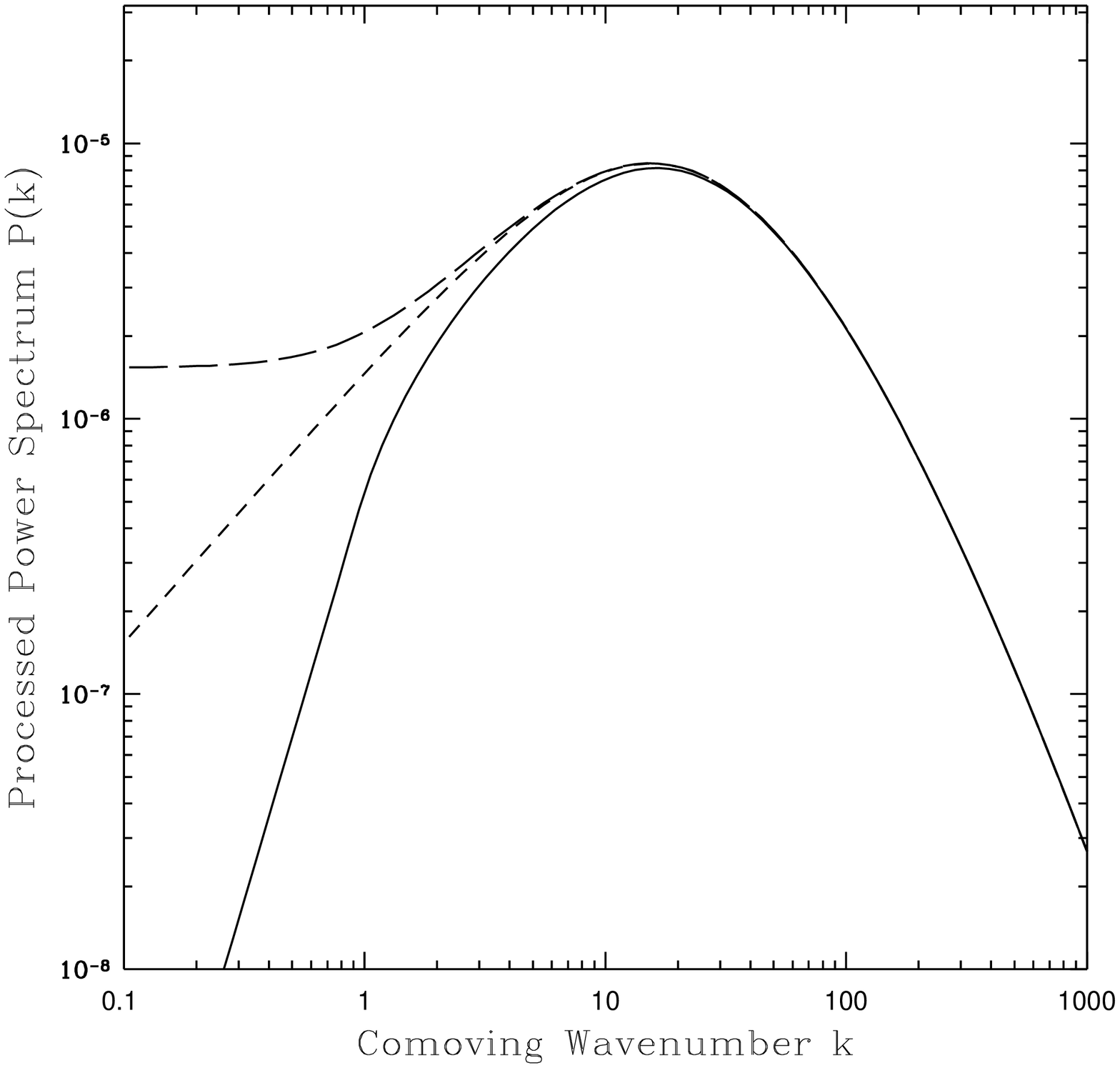}
\caption{Processed power spectra for primordial spectra that
are power laws in volume (solid curve),
wavelength (short-dash curve), and
eigenvalue of the Laplace operator
(long-dash curve).  In all cases, the
power-law index is $n=1$, and $k$ is units such that $k=1$
corresponds to the curvature scale.  
(Figure {}from Kamionkowski and Spergel 1993.)}
\label{fig: Power Spectrum}
\end{figure}

\begin{figure}[htb]
\epsfxsize=3in \epsfbox{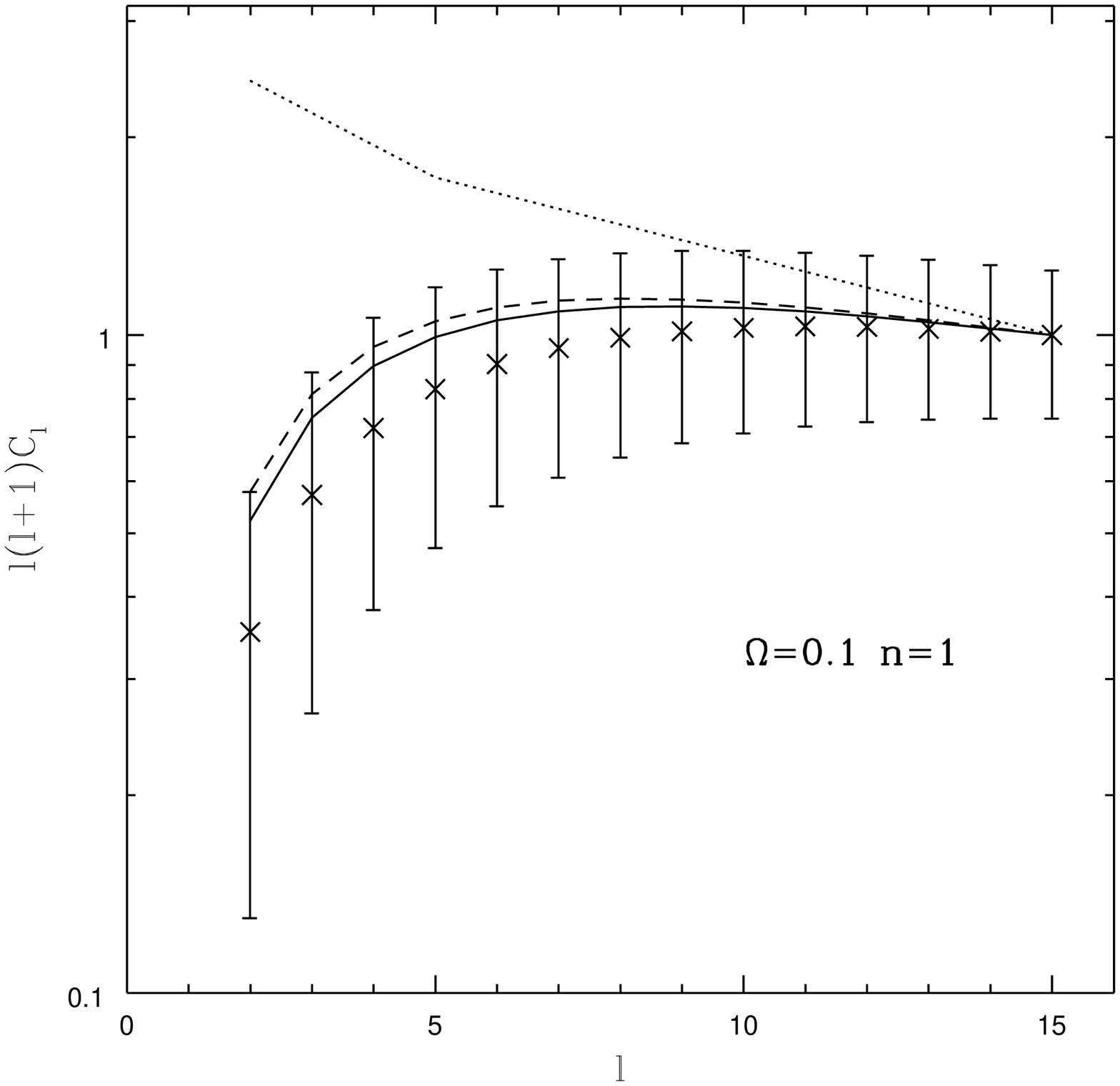}
\caption{The predicted spectrum of CMB
multipole moments, with arbitrary normalization,
for adiabatic perturbations with an $n=1$
primordial spectrum that is a power law in
volume.  The error bars are the theoretical
uncertainties due to cosmic variance.  Also
shown are the results for primordial spectra
that are $n=1$ power laws in wavenumber $k$
(solid curve) and eigenvalue $q=(k^2+1)^{1/2}$
(broken curve).  (From Kamionkowski 
and Spergel 1993.)}
The dotted line is the multipole
spectrum in a flat string-dominated universe.
\label{fig: Multipole Spectrum}
\end{figure}

	For models with scale-invariant fluctuations, the current
observations place restrictive limits on the allowed value
of $\Omega$.  Observations of large-scale structure (Vogeley et al. 1992;
Fisher et al. 1993; Baugh and Efstathiou 1993)
imply that $\Omega h \sim 0.2 - 0.3$ for a CDM-dominated universe.
Estimates of globular-cluster ages imply that the age of the
Universe should exceed 13 Gyr (Deliyannis et al. 1989).
Combining these constraints
with COBE normalization (Smoot et al. 1992) 
and the requirement that amplitude
of density fluctuations on the $8/h$ Mpc scale exceeds
0.5 restricts these models to a narrow region of parameter
space.  Figure 3 shows that these three constraints can
only be satisfied for $\Omega > 0.35$ and that large values
of $\Omega$ require that the Hubble constant is less than 50 km/s/Mpc.
Unfortunately, the COBE observations alone can not determine the
value of $\Omega$ as 
the predicted COBE multipole spectrum in an open universe with curvature
fluctuations is too
similar to the predicted flat-universe multipole spectrum: $c_l l (l+1)$
is nearly constant.
However, the implied interpretation of the COBE results is
quite different in an open-universe model.
If the Universe is open, the dominant 
source of fluctuations in the COBE experiment is not variations
in the gravitational potential at the surface of last scatter,
but rather variations in the gravitational potential at $z \sim 1/\Omega$.
This contrasts with a flat adiabatic model in which the dominant
source of fluctuations in the COBE experiment are variations
in the gravitational potential at the surface of last scatter, and
the gravitational potential is independent of time in the linear regime.
Thus if the Universe is open, COBE may be seeing microwave fluctuations
due to large scale structure at the redshifts of the most
distant quasars, rather than looking back to $z \sim 1000.$  

\begin{figure}[htb]
\epsfxsize=3in \epsfbox{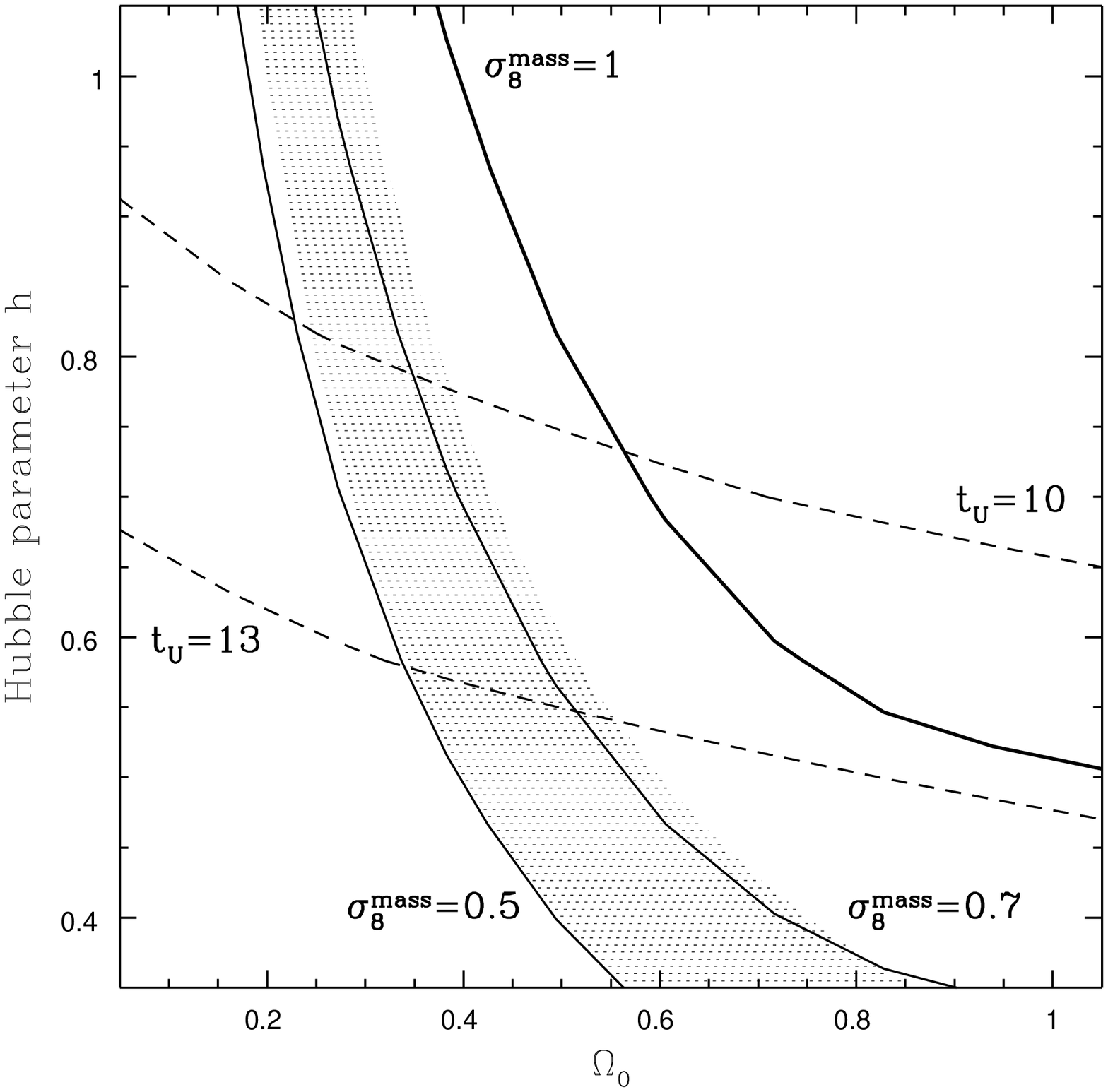}
\caption{Contour plot of $\sigma_8^{\rm mass}$ obtained
{}from COBE normalization in the $\Omega$-$h$
plane.  The heavy solid curve is the contour of
of $\sigma_8^{\rm mass}=1$; the upper and lower
lighter solid curves are contours of
$\sigma_8^{\rm mass}=0.7$ and 0.5, respectively.
The upper and lower broken curves are contours
of age of the Universe of 10 and 13 Gyrs
respectively.  The shaded region is that where
$0.2\leq\Omega h\leq0.3$, as suggested by the
observed power on large scales.
(From Kamionkowski and Spergel 1993)}
\label{fig: Allowed Model Parameters}
\end{figure}

	On scales smaller than a few degrees, ``Doppler fluctuations" are
an important additional source of CMB fluctuations (see e.g.,
Bond et al. 1993).  This process
is important for the many experiments probing the CMB on
sub-COBE angular scales.  On sub-horizon scales
at the redshift of recombination, photons, baryons and dark matter particles
are not moving at the same velocity.  The photons scatter off of the moving
electrons and this produces CMB fluctuations.  Since this process is
only important on sub-horizon scales, it produces an additional ``bump"
in the CMB angular fluctuation spectrum 
on an angular scale comparable to that subtended
by the horizon at the redshift of last scatter.  In a flat model
(with or without cosmological constant), this process produces CMB
fluctuations with a coherence scale of about $1^o$.  In open models,
this coherence scale is smaller: $\sim \Omega^{1/2}$ degrees
(Kamionkowski, Sugiyama, and Spergel 1994).  

Figure 4 shows the
COBE-normalized CMB spectrum for several values of $\Omega$ with
scale-invariant spectra of primordial adiabatic perturbations,
$\Omega_b=0.06$ and $h=0.5$.  Shown are
models with no reionization (optical depth to the surface of
last scattering $\tau=0$) and with reionization ($\tau=1$).
It was argued that although the amplitude of the Doppler peak
changes with $\tau$ (or alternative, with $\Omega_b$ and/or
$h$), the location of the peak depends primarily on the value of
$\Omega$ and is relatively insensitive to the other unknown
cosmological parameters.  Our analysis concluded
that open models are consistent with current experiments, and that
their distinctive signature is potentially detectable
with the current generation of experiments.

\begin{figure}[htb]
\epsfxsize=3in \epsfbox{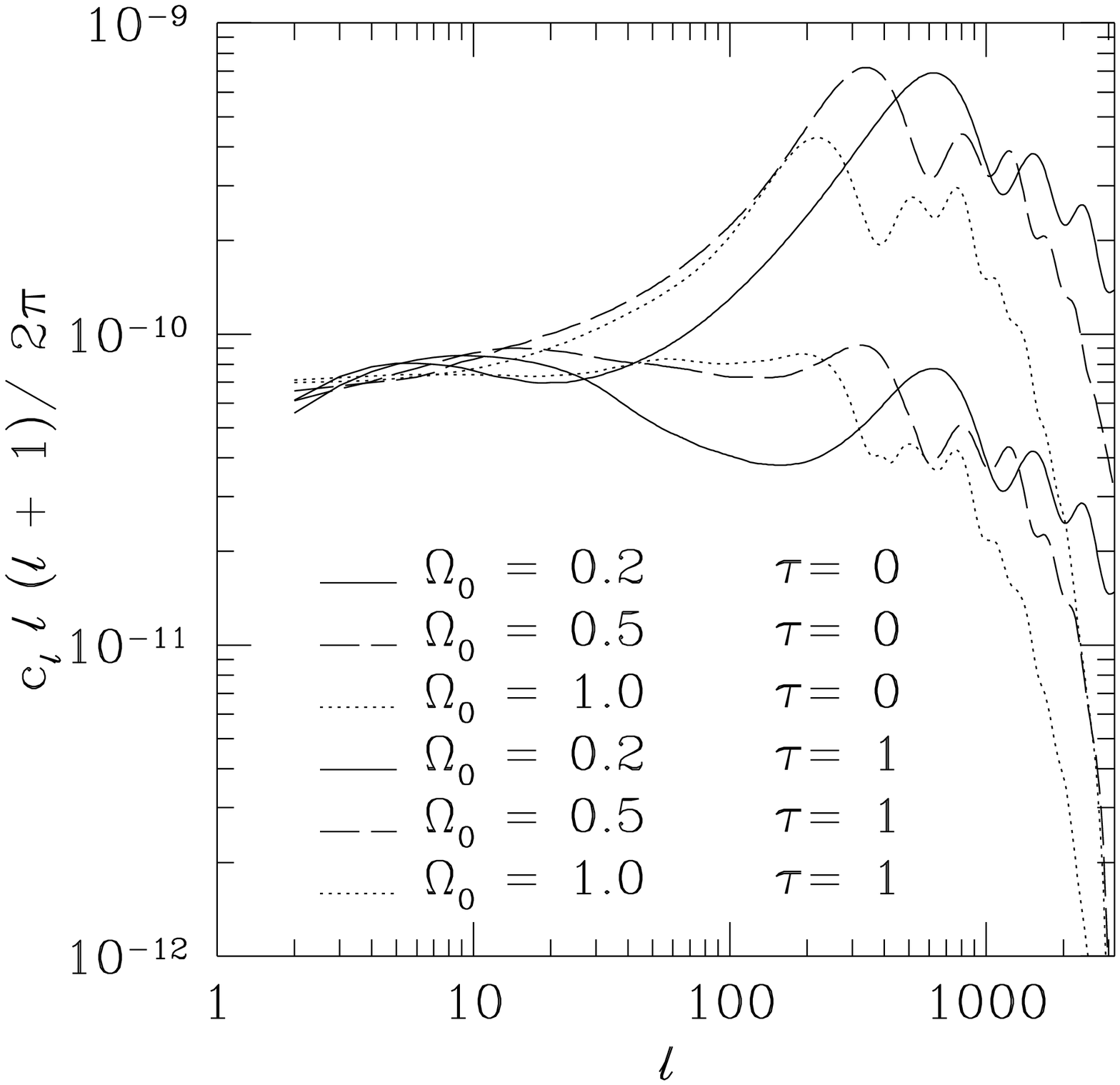}
\caption{The COBE-normalized CMB spectrum as a
function of multipole moment $l$ for
several values of $\Omega$ and for optical
depths $\tau=0$ (no reionization) and
$\tau=1$.  Here we have taken
$\Omega_b=0.06$ and $h=0.5$.
(From Kamionkowski, Spergel, and Sugiyama 1993.)}
\label{fig: Multipole Spectrum for Models with Curvature Fluctuations}
\end{figure}

	While curvature fluctuations in an open universe appear to be
compatible with a host of observations,  we still lack a natural
mechanism for generating scale-invariant fluctuations in an open-universe
model.  While there has been some work on inflationary open
universe models (Lyth and Stewart 1990; Ratra and Peebles 1994;
Kamionkowski and Liddle 1994) at this point they have not been
studied in nearly as much detail as flat models.  It should be
noted that even if The Universe is open, it may have undergone a
period of inflation which, for some reason, ended abruptly.
Although it is not clear whether such a period of ``frustrated''
inflation could solve the horizon problem, it could indeed
provide a causal mechanism for producing primordial adiabatic
density perturbations.

\section{Equation-of-State Fluctuations}

    Perhaps, the most elegant way of generating density
fluctuations in an open universe is to assume that at the start of
the FRW phase the Universe was completely homogeneous in both
density and composition and that subsequent causal physics
generated the density fluctuations that we see today.  The
generation of topological defects through symmetry breaking is a
natural way of realizing this idea (Kibble 1976). Symmetry 
breaking is a fundamental part of modern physics and most
unification schemes predict multiple phase transitions that may
have generated topological defects such as strings, monopoles, and
textures (Vilenkin 1985). It has always been difficult to reconcile
inflationary models with defects, thus, one approach is to abandon
inflation and its prediction of a flat universe and to consider
defects in an open universe (Spergel 1993). Detailed studies of these models,
however, have focused on the evolution of these defects in a flat
universe despite the difficulties in making defects compatible
with inflation.

	Defects in a flat universe have a relatively simple scaling
evolution. In the case of cosmic strings, the curvature scale of
the strings appears in numerical simulations to be a constant
fraction of the horizon size (e.g., Albrecht and Stebbins 1992;
Perivolaropoulos 1993)
In numerical simulations of global field evolution  in an expanding
universe,  the characteristic scale of textures and the spacing
between textures is also found to grow linearly with conformal time
in a radiation and matter-dominated universe (Spergel et al. 1990;
Nagasawa and Sato 1992). This scaling form has been used in many
studies of these models in flat universes (Vilenkin 1985;
Gooding et al. 1991).

	Scaling behavior can be seen analytically in the non-linear sigma
model (Turok and Spergel 1991) in the
large-N limit.  This model describes the evolution of 
an N component scalar field $\vec \psi$ constrained to lie on an
N-1 dimensional manifold: $|\vec \psi| = \psi_0$. With this
constraint, the evolution of the scalar field is described by the
non-linear sigma model (Turok 1989). 
This model is an accurate description of theories such as texture
on scales larger than the texture unwinding scale, $1/\psi_0$.
Spergel (1993) showed that in the limit of $\Omega \to 0$, the 
coherence scale grows as $\eta^{1/2}$.
Thus, during the matter and radiation dominated epochs, the scale
of symmetry breaking grows in conformal coordinates as $\eta$,
while in the curvature-dominated epoch, the scale of symmetry
breaking grows as $\eta^{1/2}$. For cosmic strings, this implies
that the string curvature scales grows with the horizon size during
the matter and radiation dominated epochs and then grows as its
square root during the curvature dominated epoch.  This is due to
the cosmic strings slowing down due to the rapid expansion of the
universe during the curvature dominated epoch.  Similar behavior is
expected for global monopoles and textures. A simple interpolation
between the  two scaling forms is 
$l_{coh}^2 =  \beta^2 a\eta/a'$.
Here, the details of the theory determine $\beta$, the
ratio of the field correlation length to the horizon size.  

	In a smooth universe, spatial variations in the pressure
in one component can lead to the generation of density
fluctuations.  It is through this process that strings, textures,
and related models produce density fluctuations.  The spatial
variations in pressure lead to a source term that produces
variations in the metric,
\begin{equation}{\cal S}_k = 4\pi G a^2 p_k\end{equation}
where $p_k$ is the harmonic expansion of the pressure in the scalar
field. Thus, calculation of the evolution of density fluctuations
and microwave fluctuations requires that we know the evolution of
the pressure in the scalar field and its auto-correlation function. 
By assuming that the pressure autocorrelation function should
depend only on the coherence scale and a constant of
proportionality set by the scale of symmetry breaking,  Spergel
(1993) was able to compute microwave and density fluctuations
semi-analytically.  This work found that the predicted CMB spectrum
was consistent with the amplitude and multipole spectrum observed
by COBE.  This motivated further investigation of defect
models in an open universe.


\begin{figure}[htb]
\epsfxsize=3in \epsfbox{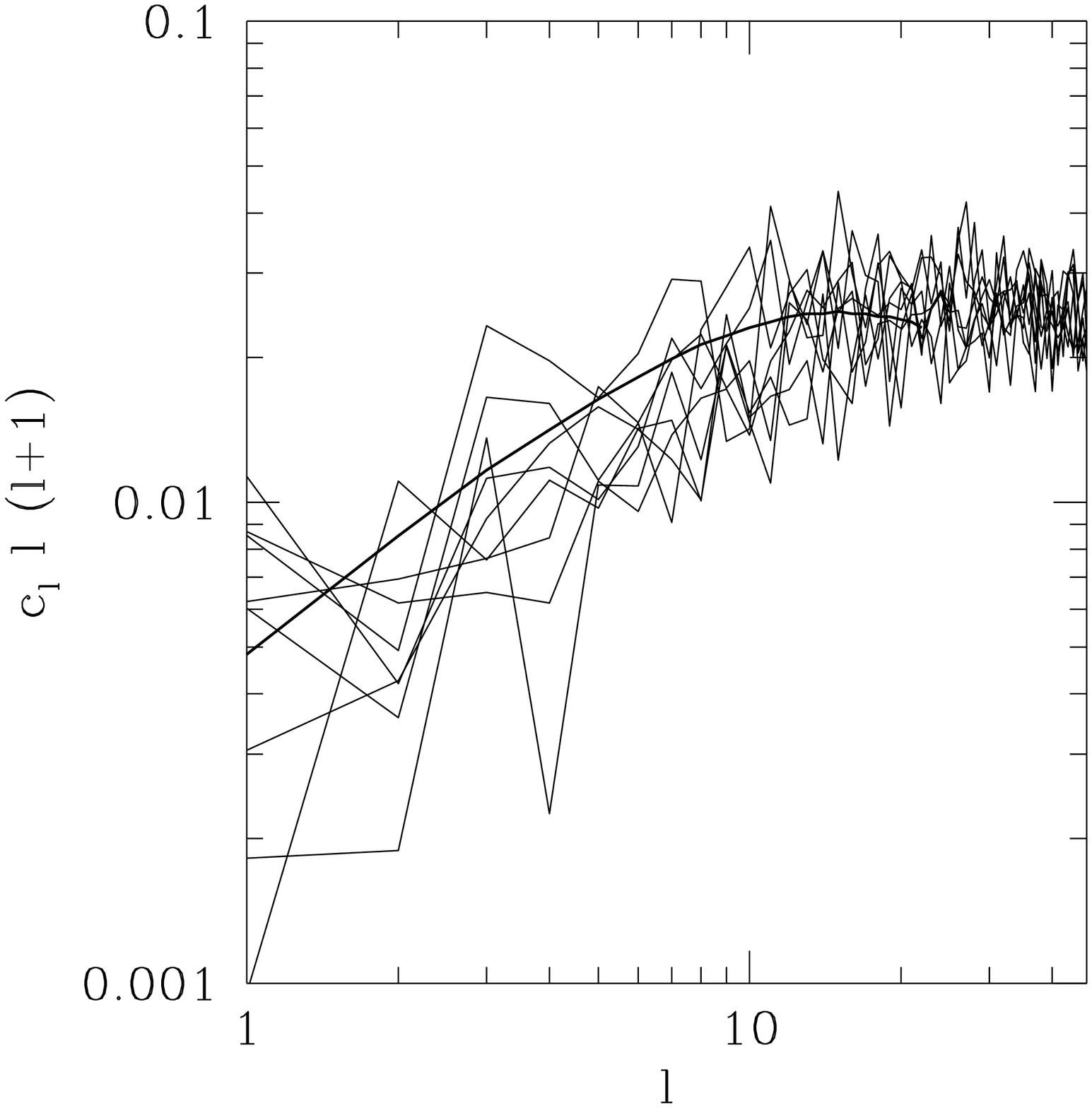}

\caption{The different solid curves trace the multipole
spectrum observed by different observers in a open
$\Omega = 0.2$ CDM dominated universe.  The temperature
fluctuations are normalized so that $8 \pi^2 G\phi_0^2 = 1$
with the tensor and vector multipole contributions scaled
{}from the flat-universe model.  The heavy line 
is the prediction of the analytic model (Spergel 1993).}
\end{figure}

Pen and Spergel (1994) investigate the texture and monopole
scenarios in detail in an open universe by using the non-linear sigma
model with the techniques described in Pen et al. (1994).  Figure
5 shows the predicted amplitude of harmonic multipoles in the
texture model in an open universe as
well as the predicted amplitudes in a scale-invariant model.  
It is intriguing that the COBE two-year spectrum (Bennett et al. 1994;
Wright et al. 1994) appears
to be better fit by the texture open universe model than
by a scale-invariant spectrum.

A very attractive cosmological model is cosmic strings in 
an open universe.  In a flat cold-dark-matter dominated universe,
cosmic-string models predict a power spectrum with a peak
at too small a physical scale (Albrecht and Stebbins 1992).
However, in an open universe, the peak of the power spectrum is shifted
to a large angular size.  As Figure 6 shows, a cosmic string
plus CDM model with $\Omega h = 0.4$ appears to be a good
fit to the galaxy fluctuations power spectrum inferred
{}from the APM survey (Baugh and Efstathiou 1993).
The gravitational-wave limits on cosmic strings are weaker
in a low-$\Omega$ universe than in a flat universe as
these limits scale as $\Omega h^2$.  Preliminary estimates
suggest that a model with cosmic strings in a cold-dark-matter
open universe is compatible with both observations of
large-scale structure, observations of CMB fluctuations and
constraints {}from the millisecond pulsar.


\begin{figure}[htb]
\epsfxsize=3in \epsfbox{newfigure2.ps}
\caption{The triangles are the power spectrum of galaxy
fluctuations computed {}from the APM survey (Baugh and Efstathiou 1993).
The lines show the predicted power spectrum for COBE-normalized
defect models with different values of $\Omega$ and $h$.
The bias is assumed to be 3 for the $\Omega =1$ and 2.5 for
the $\Omega = 0.3$ and $\Omega = 0.5$ models.  These bias
factors have a ``1 $\sigma$" uncertainty of  20\%. (From Pen and
Spergel 1994.)}
\label{fig:Comparision of COBE normalized theories to APM
Galaxy Survey}
\end{figure}

	In these equation-of-state models, the amplitude
of the low multipole moments are suppressed relative
to both flat and open curvature dominated models.  Since
there has not been any time to generate potential fluctuations
on large scales in these models, CMB fluctuations are  suppressed
on large angular scales.  Pen and Spergel (1994) find
that the multipole spectrum
for defect models rises as $l$ for $l < l_{curv}$ and then
is flat for $l > l_{curv}$, where $\pi/l_{curv}$ is
the angular size subtended by the defect coherence length
at $z \sim 1/\Omega$.  For small $l$ (large angular size),
this spectrum shape is due to incoherent contributions of different causally
disconnected regions.  For large $l$, the spectral shape reflects
that at the large redshifts at which the fluctuations on small
angular separations were generated, $\Omega$ was close to 1.

	On small angular scales (large $l$), the predictions of
topological-defect models are sensitive to assumptions
about the ionization history of the Universe.  As we lack any real
understanding of the first generation of stars, we can not with
any confidence compute the ionization history of a given cosmological model.
Fortunately, CMB observations on small angular scales should eventually
be able to constrain this history.  However, until the observations
are more certain, we must explore different assumed ionization histories.
Coulson et al. (1993) have investigated the CMB
fluctuations produced by defects in a flat fully ionized universe.  They find
on small angular scales, a fluctuation spectrum that is flat
on scales larger than the defect coherence
size at the surface of last scatter and suppressed for
larger $l$.  Since they assume that the Universe
was fully ionized, this implies that CMB fluctuations are suppressed
at $l > 60$ for cosmic-string, monopole, and texture models.  If the
universe is open, then the angular scale associated with the surface
of last scatter is smaller and the Coulson et al. (1993) results have to be
rescaled so that fluctuations are flat between $l \sim l_{curv}$ and 
$l \sim 60/\Omega^{1/2}$.  Thus, if the Universe was
fully ionized, the coherence scale of the CMB fluctuations
should be the coherence scale of the defect at the
surface of last scatter.  On the other hand, the Universe may not
have been reionized at large redshift by an early generation of stars.
Bouchet, Bennett, and Stebbins (1988) 
assume no reionization in their investigation
of string models in a flat universe and predict a much higher amplitude of
CMB fluctuations on small angular scales.  

     A potentially distinctive signature of topological-defect
models is that  the distribution of temperature fluctuations are
expected to be non-Gaussian.  This is in marked contrast to the
inflationary scenario that predicts a flat universe and Gaussian
microwave fluctuations.  
Coulson et al. (1993) suggest that the
non-Gaussian fluctuations should be largest on small angular
scales and most apparent in monopole and texture models.  
It is intriguing that the low level of fluctuations found
by the ACME/HEMT experiment (Gaier et al. 1992, Schuster et al. 1993)
in one region of
the sky, $(\delta T/T)_{rms}(1^o) < 1.2\times 10^{-5}$,  is
difficult to reconcile with the detections by the MAX experiment
(Gundersen et al. 1993)
and the MSAM experiment  (Cheng et al. 1993)
in other regions of the sky in a Gaussian
model.

\section{Conclusions}

	Current observations of the CMB fluctuation spectrum are compatible
with both open and flat geometries.  However, future observations
may be able to detect characteristic signatures of open-universe
models.  If the density fluctuations that formed
galaxies were primordial and the Universe
is open, then the location of the Doppler peak is determined by
the geometry of the Universe.  On the other hand, open-universe models
in which the density fluctuations
were produced by topological defects at low redshifts do not
have a strong Doppler peak, but they do have a
characteristic signature:
a dramatic suppression of fluctuations on large angular scales.  
While there is  a weak suppresion in the curvature models,
this effects is more dramatic in the defect models.
These defect models
also predict non-Gaussian CMB fluctuations.  Future experiments
that measure the CMB spectrum can potentially determine the geometry
of the Universe, an observation that would have significant implications
for our understanding of the origin of the Universe and its ultimate
fate.

\section{References}
\vspace{1pc}

\newcommand{\re}{}

\re
Abbott, L. F. and Schaefer, R. K. 1986, {\it Ap. J.} {\bf 308}, 546.

\re
Albrecht, A. and Stebbins, A. 1992, {\it Phys. Rev. Lett.}
{\bf 68}, 2121.

\re
Baugh, C. and Efstathiou, G. 1993, {\it M.N.R.A.S.} {\bf 265}, 145.

\re
Bennett, C. et al. 1994, COBE preprint 1994-1.

\re
Bouchet, F. R., Bennett, D. P., and Stebbins, A. 1988, {\it
Nature} {\bf 335}, 410.

\re
Bond, J. R., Crittenden, R., Davis, R. L., Efstathiou, G.,
and Steinhardt, P. J. 1993, CITA preprint.

\re
Cheng, E.S. et al. 1993, Princeton University Physics preprint.

\re
Coulson, D., Ferreira, P., Graham, P. and Turok, N. 1993,
Princeton University Physics preprint.

\re
Deliyannis, C. P., Demarque, P., and Pinsonneault, M. H. 1989, {\it
Ap. J.} {\bf 347}, L73.

\re
Fisher, K. B., Davis, M., Strauss, M. A., Yahil, A., and
Huchra, J. P. 1993, {\it Ap. J.} {\bf 402}, 42.

\re
Gaier, T., Schuster, J., Gundersen, J., Koch, T., Seiffert. M.,
Meinhold, P., and Lubin, P. 1992, {\it Ap. J.} {\bf 398,} L1.

\re
Gooding, A., Spergel, D. N., and Turok, N. 1991, {\it Ap. J. (Letters)}
{\bf 372}, L5.

\re
Gundersen, J. O. et al. 1993, {\it Ap. J. (Letters)} {\bf 413,} L1.

\re
Kamionkowski, M. and Liddle, A. R. 1994, work in progress.

\re
Kamionkowski, M. and Spergel, D. N. 1994, submitted to {\it Ap. J.}

\re
Kamionkowski, M., Spergel, D. N., and Sugiyama, N. 1994, submitted to
{\it Ap. J. (Letters)}.

\re
Kibble, T. W. B. 1976, {\it J. Phys} {\bf A9}, 1387.

\re
Kodama, H. and Sasaki, M. 1984, {\it Prog. of Theor. Physics}
{\bf 78},  1.

\re
Pen, U., Spergel, D. N., and Turok, N. 1994, to appear in {\it
Phys. Rev. D}.

\re
Pen, U. and Spergel, D.N. 1994, in preparation.

\re
Perivolaropoulos, L. 1993, {\it Phys. Lett.} {\bf B298}, 305.

\re
Ratra, B. and Peebles, P. J. E. 1993, private communication.

\re
Schuster, J. et al. 1993, {\it Ap. J. (Letters)} {\bf 412,} L47. 

\re
Smoot, G. F. et al. 1992 {\it Ap. J.} 396, L1. 

\re
Spergel, D. N., Turok, N., Press, W. H. and Ryden, B. S. 1990,
{\it Phys. Rev. D} {\bf 43}, 1038.

\re
Spergel, D. N. 1993, {\it Ap. J. (Letters)} {\bf 412} L12.

\re
Sugiyama, N. and Gouda, N. 1992, {\it Prog. Theor. Phys.} {\bf 88}, 803.

\re
Turok, N. 1989, {\it Phys. Rev. Lett.} {\bf 63}, 2625.

\re
Turok, N. and Spergel, D. N. 1991, {\it Phys. Rev. Lett.}
{\bf 66}, 24.

\re
Vilenkin, A. 1985, {\it Phys. Rep.}, {\bf 121}, 263.

\re
Vogeley, M. S., Park, C., Geller, M. J. and Huchra, J. P. 1992,
{\it  Ap. J.} 391, L5.

\re
White, S. D. M. 1992, in {\it Clusters and Superclusters of
Galaxies,} ed. A.C. Fabian (Netherlands, Kluwer Academic
Publishers), 17.

\re
Wilson, M. L. 1983, {\it Ap. J.} {\bf 273}, 2.

\re
Wright, E., et al. 1994, COBE preprint 1994-2.

\end{document}